%% file: main.tex
\documentclass[10pt,conference]{IEEEtran}
\input{meta/packages.tex}
\input{meta/macros.tex}

\input{meta/findings.tex}

\begin{document}

\title{Predicting Merge Conflicts in Collaborative Software Development}

\author{\IEEEauthorblockN{Moein Owhadi-Kareshk}
\IEEEauthorblockA{
\textit{University of Alberta, AB, Canada}\\
owhadika@ualberta.ca}
\and
\IEEEauthorblockN{Sarah Nadi}
\IEEEauthorblockA{
\textit{University of Alberta, AB, Canada}\\
nadi@ualberta.ca}
\and
\IEEEauthorblockN{Julia Rubin}
\IEEEauthorblockA{
\textit{University of British Columbia, BC, Canada}\\
mjulia@ece.ubc.ca}
}

\IEEEoverridecommandlockouts
\IEEEpubid{\makebox[\columnwidth]{978-1-7281-2968-6/19/\$31.00~\copyright2019 IEEE \hfill} \hspace{\columnsep}\makebox[\columnwidth]{ }}

\maketitle

\IEEEpubidadjcol

\input{sections/abstract.tex}

\input{sections/introduction.tex}
\input{sections/related.tex}
\input{sections/methodology.tex}
\input{sections/evaluation_setup.tex}

\input{sections/results.tex}

\input{sections/threats.tex}
\input{sections/conclusion.tex}
\input{sections/acknowledgement.tex}

\bibliographystyle{IEEEtran}
\balance
\bibliography{references/references}

\end{document}

%% file: meta/packages.tex
\usepackage[utf8]{inputenc}
\usepackage{amsmath}
\usepackage{amsfonts}
\usepackage{amssymb}
\usepackage{cite}
\usepackage{url}
\usepackage[table]{xcolor}
\usepackage{graphicx}
\usepackage{xspace}
\usepackage{enumitem}
\usepackage{balance}

\usepackage{multirow}
\usepackage{tabularx}

%% file: meta/macros.tex
\definecolor{darkgreen}{HTML}{3C8031}
\definecolor{jcolor}{rgb}{0.5,0,0.5}

\newcolumntype{L}{>{\raggedright\arraybackslash}X}

\newcommand{\initreposNumperlang}{$150$\xspace}
\newcommand{\initreposNum}{$1,050$\xspace}
\newcommand{\reposNum}{$744$\xspace}

\newcommand{\mergeScenarioNum}{$267,657$\xspace}

\newcommand{\mergeScenarioNumConflicting}{$21,734$\xspace}
\newcommand{\confRate}{$8.12\%$\xspace}
\newcommand{\confRateWO}{$0.0812$\xspace}

\newcommand{\featureSetNum}{9\xspace}
\newcommand{\featureNum}{28\xspace}

\newcommand{\predictionTime}{$0.1$ seconds\xspace}
\newcommand{\predictionTimeSTD}{$0.02$ seconds\xspace}

\newcommand{\git}{Git\xspace}
\newcommand{\github}{GitHub\xspace}
\newcommand{\textual}{textual conflicts\xspace}

\newcommand{\progLangs}{C, C++, C\#, Java, PHP, Python, and Ruby\xspace}
\newcommand{\progLangNum}{seven\xspace}

\newcommand{\fNCMin}{$0.95$\xspace}
\newcommand{\fNCMax}{$0.97$\xspace}
\newcommand{\fCMin}{$0.57$\xspace}
\newcommand{\fCMax}{$0.68$\xspace}

\newcommand{\pNCMin}{$0.97$\xspace}
\newcommand{\pNCMax}{$0.98$\xspace}
\newcommand{\pCMin}{$0.48$\xspace}
\newcommand{\pCMax}{$0.63$\xspace}

\newcommand{\rNCMin}{$0.93$\xspace}
\newcommand{\rNCMax}{$0.96$\xspace}
\newcommand{\rCMin}{$0.68$\xspace}
\newcommand{\rCMax}{$0.83$\xspace}

%% file: meta/findings.tex
\usepackage[absolute,overlay]{textpos}
\usepackage{tikz}
\usepackage{xcolor}
\usepackage{calc}

\newcounter{implicationctr}

\newcommand{\finding}[2]{}

\newcommand{\implication}[1]{\refstepcounter{implicationctr}\emph{Implication \theimplicationctr:}\label{implication:#1}}

\newlength{\boxw}
\newlength{\boxh}
\newlength{\shadowsize}
\newlength{\boxroundness}
\newlength{\tmpa}
\newsavebox{\shadowblockbox}

\newenvironment{findingenv}[2]%
{\vspace{0.2cm}\noindent
\begin{lrbox}{
\shadowblockbox
}
\begin{minipage}{.98\columnwidth}
\finding{#1}{#2}}%
{\end{minipage}\end{lrbox}%
\settowidth{\boxw}{\usebox{\shadowblockbox}}%
\settodepth{\tmpa}{\usebox{\shadowblockbox}}%
\settoheight{\boxh}{\usebox{\shadowblockbox}}%
\addtolength{\boxh}{\tmpa}%
\begin{tikzpicture}
\addtolength{\boxw}{\boxroundness * 2}
\addtolength{\boxh}{\boxroundness * 2}

\foreach \x in {0,.05,...,1}
{
\setlength{\tmpa}{\shadowsize * \real{\x}}
\fill[xshift=\shadowsize - 1pt,yshift=-\shadowsize + 
1pt,black,opacity=.04,rounded corners=\boxroundness] 
(\tmpa, \tmpa) rectangle +(\boxw - \tmpa - \tmpa, \boxh - \tmpa - 
\tmpa);
}

\filldraw[fill=white!50, draw=black!80, rounded corners=\boxroundness] (0, 
0) rectangle (\boxw, \boxh);
\draw node[xshift=\boxroundness,yshift=\boxroundness,inner sep=0pt,outer 
sep=0pt,anchor=south west] (0,0) {\usebox{\shadowblockbox}};
\end{tikzpicture}\vspace{0cm}%
}

{\vspace{0.2cm}\noindent
\begin{lrbox}{
\shadowblockbox
}
\begin{minipage}{.98\columnwidth}
\implication{#1}}%
{\end{minipage}\end{lrbox}%
\settowidth{\boxw}{\usebox{\shadowblockbox}}%
\settodepth{\tmpa}{\usebox{\shadowblockbox}}%
\settoheight{\boxh}{\usebox{\shadowblockbox}}%
\addtolength{\boxh}{\tmpa}%
\begin{tikzpicture}
\addtolength{\boxw}{\boxroundness * 2}
\addtolength{\boxh}{\boxroundness * 2}

\foreach \x in {0,.05,...,1}
{
\setlength{\tmpa}{\shadowsize * \real{\x}}
\fill[xshift=\shadowsize - 1pt,yshift=-\shadowsize + 
1pt,black,opacity=.04,rounded corners=\boxroundness] 
(\tmpa, \tmpa) rectangle +(\boxw - \tmpa - \tmpa, \boxh - \tmpa - 
\tmpa);
}

\filldraw[fill=white!50, draw=black!80, rounded corners=\boxroundness] (0, 
0) rectangle (\boxw, \boxh);
\draw node[xshift=\boxroundness,yshift=\boxroundness,inner sep=0pt,outer 
sep=0pt,anchor=south west] (0,0) {\usebox{\shadowblockbox}};
\end{tikzpicture}\vspace{0cm}%
}

%% file: sections/abstract.tex
\begin{abstract}
\textit{Background.} During collaborative software development, developers often use branches to add features or fix bugs. 
When merging changes from two branches, conflicts may occur if the changes are inconsistent.
Developers need to resolve these conflicts before completing the merge, which is an error-prone and time-consuming process.
Early detection of merge conflicts, which warns developers about resolving conflicts before they become large and complicated, is among the ways of dealing with this problem.
Existing techniques do this by continuously pulling and merging all combinations of branches in the background to notify developers as soon as a conflict occurs, which is a computationally expensive process. One potential way for reducing this cost is to use a machine-learning based conflict predictor that filters out the merge scenarios that are not likely to have conflicts, i.e. \textit{safe merge scenarios}.

\textit{Aims.}  In this paper, we assess if conflict prediction is feasible.  

\textit{Method.}
We design a classifier for predicting merge conflicts, based on \featureSetNum light-weight \git feature sets.
To evaluate our predictor, we perform a large-scale study on \mergeScenarioNum merge scenarios from \reposNum \github repositories in \progLangNum programming languages. 

\textit{Results.} Our results show that we achieve high f1-scores, varying from \fNCMin to \fNCMax for different programming languages, when predicting safe merge scenarios. The f1-score is between \fCMin and \fCMax for the conflicting merge scenarios. 

\textit{Conclusions.} Predicting merge conflicts is feasible in practice, especially in the context of predicting safe merge scenarios as a pre-filtering step for speculative merging.
\end{abstract}


%% file: sections/introduction.tex
\section{Introduction}\label{sec:intro}

Modern software systems are commonly built by a large, distributed teams of developers.
Thus, improving the collaborative software development experience is important.
Distributed Version Control Systems (VCSs), such as \git, and social coding platforms, such as \github, have made such collaborative software development easier.
However, despite its advantages, collaborative software development also gives rise to several issues~\cite{bird2009promises,kalliamvakou2014promises}, including merging and integration problems\cite{mckee2017software,accioly2018understanding}.

When two developers change the same part of the code simultaneously, \git cannot decide which change to choose and reports \textit{\textual}. In this situation, the developers need to resolve the conflict manually, which is an error-prone and time-consuming task that wastes resources~\cite{sarma2012palantir,bird2012assessing}. 

Given the cost of merge conflicts and integration problems, many research efforts have advocated earlier resolution of conflicts~\cite{brun2013early,guimaraes2012improving,sarma2012palantir}. Previous work has shown that lack of awareness of changes being done by other developers can cause conflicts~\cite{estler2014awareness}, and since infrequent merging can decrease awareness, it increases the chance of conflicts. 
To address that, \textit{proactive merge-conflict detection} warns developers about possible conflicts before they actually attempt to merge, i.e., before they try to push their changes or pull new changes. 
With proactive conflict detection, developers get warned early about conflicts so they can resolve them soon instead of waiting till later when they get large and complicated.

In the literature, proactive conflict detection is typically based on \textit{speculative merging}~\cite{baumgartner2011incremental,guimaraes2012improving,brun2011proactive,kasi2013cassandra}, where all combinations of available branches are pulled and merged in the background. While a single textual merge operation is cheap, constantly pulling and merging a large number of branch combinations can quickly get prohibitively expensive. One opportunity we foresee for decreasing this cost is to avoid performing speculative merging on \textit{safe merge scenarios} that are unlikely to have conflicts.
To accomplish this, we can leverage machine learning techniques to design a classifier for predicting merge conflicts. The question is whether such a classifier works well in practice.

To the best of our knowledge, there have been two attempts at predicting merge conflicts in the past~\cite{lessenich2018indicators,accioly2018analyzing}. 
The first study~\cite{lessenich2018indicators} looked for correlations between various features and merge conflicts and found that none of the features have a strong correlation with merge conflicts.
The authors concluded that merge conflict prediction may not be possible.
However, we argue that lacking correlation does not necessarily preclude a successful classifier, especially since the study did not consider the fact that the frequency of conflicts is low in practice and most of the standard form of statistics and machine learning techniques cannot handle imbalanced data well. 
The second study~\cite{accioly2018analyzing} investigates the relationship between two types of code changes, edits to the same method and edits to dependent methods, and merge conflicts.
The authors report recall of $82.67\%$ and precision of $57.99\%$ based on counting how often a merge scenario that had the given change was conflicting.
This means that this second study does not build a prediction model that is trained on one set of data and evaluated on unseen merge scenarios.

Since neither of the above work built a prediction model that is suitable for imbalanced data and has been tested on unseen data, it is still not clear if predicting merge conflicts is feasible in practice, especially while using features that are not computationally expensive to extract. 
In this paper, we investigate if merge conflicts can be predicted using \git features, i.e. information that can be inexpensively extracted via \git commands. Specifically, we focus on the following two research questions:

\begin{itemize}
\item \textit{\textbf{RQ1:}} Which characteristics of merge scenarios have more impact on conflicts?
\item \textit{\textbf{RQ2:}} Are merge conflicts predictable using only git features?
\end{itemize}

To answer these questions, we study \reposNum well-engineered repositories that are listed in the reaper dataset~\cite{munaiah2017curating}, and that are written in 7 different programming languages (\progLangs). We collect \mergeScenarioNum merge scenarios from these repositories and design a separate classifier for the repositories in each programming language.
To design our classifiers, we use a total of nine feature sets that can be extracted solely through \git commands. We intentionally use only features that can be extracted from version control history so that our prediction process can be efficient (e.g., as opposed to features that may require code analysis). Furthermore, we use Decision Tree~\cite{quinlan1986induction} and ensemble machine learning techniques, specifically Random Forest~\cite{liaw2002classification}, to take into account the specific characteristics of merge data, such as being imbalanced, in our classifiers.  
To the best of our knowledge, this work presents the largest merge-conflict prediction study to date. We have almost $13$ times the number of merge scenarios used by recent work~\cite{lessenich2018indicators}, and our work is the first that is evaluated with repositories written in several programming languages.
We publish all our code and evaluation data in an online artifact page~\cite{artifact}.


Despite confirming the lack of significant correlation found by previous work~\cite{lessenich2018indicators}, our prediction results show that lack of strong correlation does not necessarily mean that a machine learning classifier would perform poorly. 
Our results show that a Random Forest classifier using all our feature sets predicts conflicting merge scenarios with a precision of \pCMin to \pCMax and a recall from \rCMin to \rCMax across the different programming languages. 
These numbers show that the predictors are capable of identifying conflicting merge scenarios, but that their performance may not be that reliable in practice.
On the other hand, our results show that the same classifiers can identify \textit{safe merges} (i.e., those without a conflict) extremely well: a precision of \pNCMin to \pNCMax, and recall between \rNCMin to \rNCMax across the different programming languages.
The above results mean that while the classifiers may not be able to precisely predict conflicts, they can predict non-conflicting scenarios with high accuracy.
This is good news for speculative merging, because when the predictor marks a merge scenario as safe, speculative merging can confidently avoid performing the merge in the background for this scenario.
This reduces the number of merges that need to be done in the background, which reduces some of the computational costs involved.

To summarize, the contributions of this paper are:

\begin{itemize}
\item We perform the largest merge-conflict prediction study to date, using \mergeScenarioNum merge scenarios extracted from \reposNum \github well-engineered repositories written in different programming languages.
\item We create a set of potential predictive features for merge conflicts based on the literature on software merging.
\item We apply systematic statistical machine learning strategies for handling the imbalance in software merging data.
\item We design effective machine learning classifiers for \textual in \progLangNum programming languages. Our classifiers can be used as a pre-filtering step in the context of speculative merging.
\end{itemize}

%% file: sections/related.tex
\section{Related Work}\label{sec:related}

In this section, we discuss the previous software merging literature that is most related to our work. We first explain different merging techniques and then describe the work that applies and analyzes these techniques in practice. Finally, we discuss proactive conflict detection, since it can potentially benefit from effective merge-conflict predictors.

\subsection{Merging Methods}
 
We first provide a brief summary of existing merging techniques. For a more comprehensive classification, we refer the reader to the Mens' survey on software merging~\cite{mens2002state}.

\git is one of the most popular VCSs~\cite{apel2011semistructured}, and \github is a social coding platform that hosts git-based projects. Git uses line-based unstructured merging, triggered through \texttt{git merge}, which is the most basic and popular merging technique~\cite{brindescu2014centralized}~\cite{apel2012structured}. 
\git is an \textit{unstructured} merging tool that is language-independent and can be employed for merging large repositories containing a variety of text files such as code, documentation, configuration, etc.
In other words, it does not consider the structure of the code (or any underlying tracked file); when the same text in a file has been simultaneously edited, \git reports a \textit{textual conflict}.

On the other hand, \textit{structured merge} tools~\cite{westfechtel1991structure}~\cite{buffenbarger1995syntactic}, e.g., FSTMerge~\cite{fstmerge}, leverage information about the underlying code structure through analyzing the corresponding Abstract Syntax Tree (AST). Since differencing a complete AST is expensive, \textit{semi-structured} merge tools, such as JDime~\cite{jdime}, improve performance by producing a partial AST that expands only until the method level, with complete method bodies in the leaves. Structured merge is then used for the main nodes of the tree, while unstructured merge is used for the method bodies in the leaves.

In this paper, we focus on \textual as reported by \git, since these are the most common types of conflicts developers face in their typical work flow. Note that when describing our work after Section~\ref{sec:related}, we often use the only the term \textit{conflict} for brevity.
 
\subsection{Empirical Studies on Software Merging}

Previous studies compared the above merge techniques in practice in terms of speed, quality of resolutions, and the complexity of reported conflicts. For example, Cavalcanti et al.~\cite{cavalcanti2017evaluating} focused on unstructured and semi-structured merge tools and found that using semi-structured merge significantly reduces the number of conflicts. The authors also found that the output of semi-structured merge is easier to understand and resolve. In a follow-up work, Accioly et al.~\cite{accioly2018understanding} investigated the structure of code changes that lead to conflicts with semi-structured tools. The study showed that in most of the conflicting merge scenarios, more than two developers are involved. Moreover, this study showed that code cloning can be a root cause of conflicts. While semi-structured merge is faster than structured merge and more precise than unstructured merge, it is still not used in software industry due to the effort that is needed in order to support new programming languages.
A recent large-scale empirical study by Ghiotto et al.~\cite{menezes2018nature} also investigated various characteristics of textual merge conflicts, such as their size and resolution types. The results suggests that since merge conflicts vary greatly in terms of their complexity and resolutions, having an automatic tool that can resolve all types of conflicts is likely not feasible.

One approach for reducing the resolution time is selecting the right developer to perform the merging based on their previous performance and changes~\cite{costa2016tipmerge}. Other work looked at specific types of changes that may affect merge conflicts. For example, Dig et al.~\cite{dig2008effective} introduced a refactoring-aware merging technique that can resolve conflicts in the presence of refactorings. A recent study also shows that 22\% of the analyzed \git conflicts involved refactoring operations in the conflicting code~\cite{MahmoudiSANER19}. 

\subsection{Proactive Conflict Detection}

There are several approaches to increase the awareness of developers by detecting conflicts early. Awareness of changes other team members may be making is a key factor in team productivity and reduces the number of conflicts~\cite{estler2014awareness}. Syde~\cite{hattori2010syde} is a tool for increasing awareness through sharing the code changes present in other developers' workspaces. Similarly, Palantir~\cite{sarma2012palantir} visually illustrates code changes and helps developers avoid conflicts by making them aware of changes in private workspaces. Crystal~\cite{brun2013early} is a visual tool that uses speculative analysis to help developers detect, manage, and prevent various types of conflicts. Cassandra~\cite{kasi2013cassandra} is another tool to minimize conflicts by optimizing task scheduling, with the goal of minimizing simultaneous edits to the same files. MergeHelper~\cite{nishimura2016supporting} helps developers find the root cause of merge conflicts by providing them with the historic edit operations that affected a given class member.

Guimar{\~a}es et al.~\cite{guimaraes2012improving} propose to continuously merge, compile, and test committed and uncommitted changes to detect conflicts as early as possible. However, such an approach is likely expensive given the large number of combinations of branches and developer changes in large projects.

Accioly et al.~\cite{accioly2018analyzing} investigate whether the occurrence of events such as \textit{edits to the same method} and \textit{edits to directly dependent methods} can be used to predict conflicts. However, they do not actually build a prediction model. Instead, they count the number of times each of the above features exists when a conflict occurs versus when the merge is successful.  Based on such counts, their results show a precision of 57.99\% and a recall of 82.67\%.

Leßenich et al.~\cite{lessenich2018indicators} investigate the correlation between various code and \git features and the likelihood of conflicts. To create a list of features they investigated, they first surveyed 41 developers. The developers mentioned seven features that can potentially cause conflicts. However, after analyzing 21,488 merge scenarios in 163 Java repositories, the authors could not find a correlation between these features and the likelihood of conflicts. We speculate that one reason for not capturing such relationships is using stepwise-regression which may not be an effective model for non-linear data, such as that collected from merge scenarios.

In this paper, we investigate merge-conflict prediction by creating a list of nine feature sets that can potentially impact conflicts. Our list is based on previous work in the areas of software merging and code review~\cite{lessenich2018indicators, kononenko2018studying, fan2018early, estler2014awareness, sarma2012palantir}. Our work is different from all the above in that we use statistical machine learning to create a classifier, for each programming language, that can predict conflicts in unseen merge scenarios. 

%% file: sections/methodology.tex
\section{Building a Merge-conflict Classifier}\label{sec:methodology}

Given a merge scenario based on two branches, our goal is to predict whether a merge conflict will occur.
In this section, we describe how we prepare the data that is needed for predicting merge conflicts, as well as how we train a classifier.  
Figure~\ref{fig:methodology} shows an overview of our methodology, which consists of three stages, as follows:

\begin{figure}[t!]
\centering
  \includegraphics[width=0.35\textwidth]{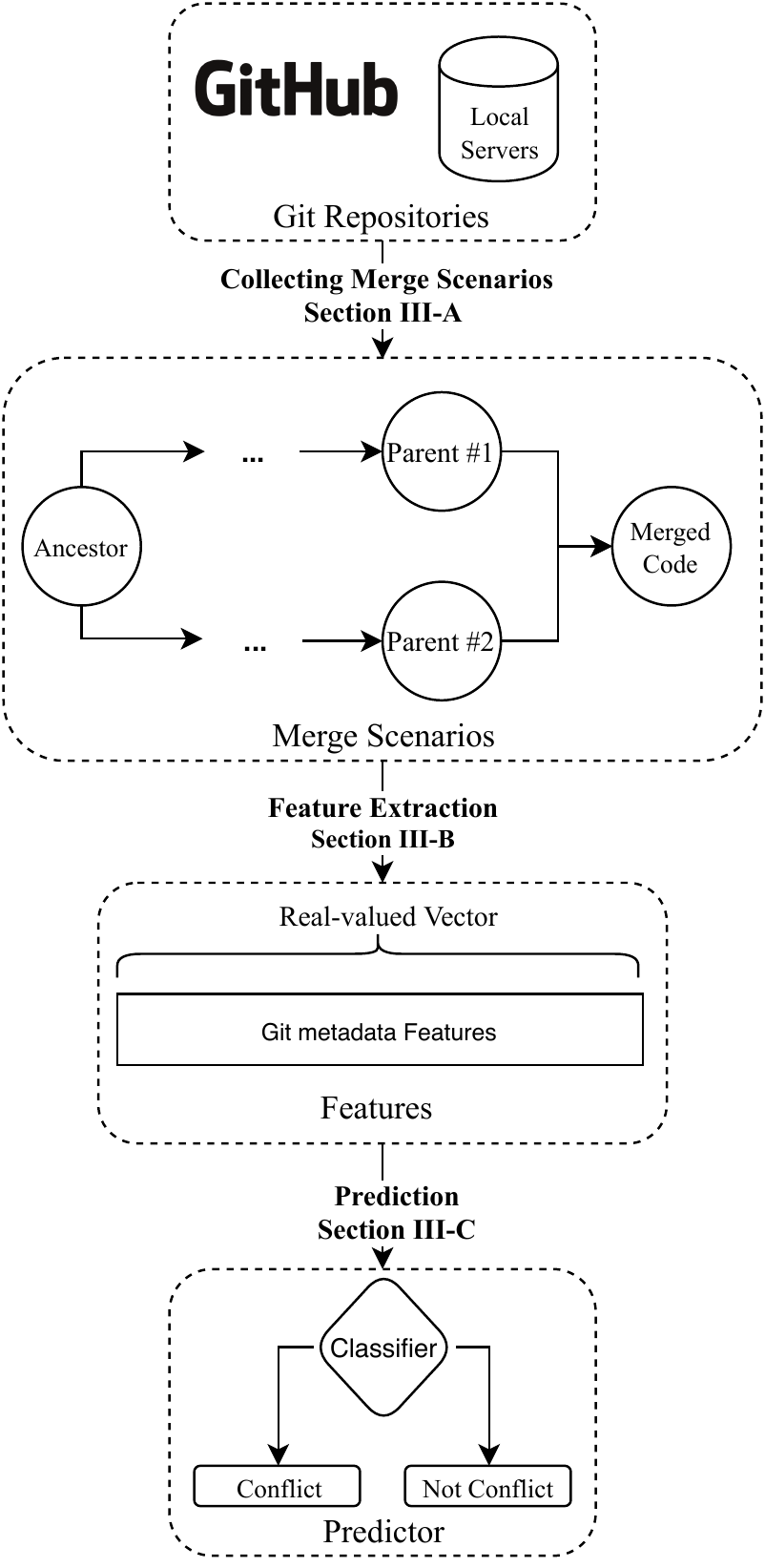}
  \caption{Methodology for Creating the Proposed Conflict Predictor}
  \label{fig:methodology}\vspace{-0.5cm}
\end{figure}

\begin{enumerate}

\item \textbf{\textit{Collecting Merge Scenarios (Section~\ref{Subsection:Merge Replaying})}} As a first step, we need to collect merge scenarios.
We do so by mining the \git history of the target repositories. 

\item \textbf{\textit{Feature Extraction (Section~\ref{Subsection:Feature Extraction}):}} In the second stage, we extract the features that we will later use to build the prediction model. Using the \git history, we extract features from both branches being merged.

\item \textbf{\textit{Prediction (Section~\ref{Subsection:Prediction}):}} In the last stage, we use statistical machine learning techniques to build a prediction model. Specifically, we use a binary classifier that aims to separate conflicting and safe merge scenarios. Since conflicts happen in only a few numbers of merge scenarios, the classifier should be capable of handling imbalanced data. 
\end{enumerate}

\subsection{Collecting Merge Scenarios} \label{Subsection:Merge Replaying}
In order to train a classifier, we need a large set of labeled merge scenarios. 
Fortunately, merge scenarios can be identified from a repository's \git history.
However, unfortunately, not all information about a merge scenario (e.g., whether there was a conflict or not) is available in \git's data.
Therefore, to identify merge scenarios and determine whether the merge resulted in a conflict, we use a \textit{replaying} method where we re-perform the merge at that point of history and record the outcome.

The input for this stage is a list of \git repositories to be analyzed.  After cloning all repositories, we use MERGANSER, an open-source toolchain we developed for extracting merge-scenario data~\cite{OwhadiKareshkMSR19}, to analyze their histories. MERGANSER considers only $3$-way merge scenarios and ignores $n$-way merges, which are called \textit{octopus merges} in \git. In $3$-way merge scenarios, the \textit{Merge Commit} has two parents (\textit{Parent \#1} and \textit{Parent \#2}) and these parents have a \textit{Common Ancestor}. MERGANSER, thus, identifies target merge scenarios by looking for commits with two parents. It then replays all identified $3$-way merge scenarios by checking out Parent \#1 and then using the \texttt{git merge} command to merge Parent \#2's changes. We use \git merge's default options, which uses the recursive merge strategy~\cite{git}. To detect conflicts after replaying, our toolchain searches for the phrase \texttt{Automatic merge failed; fix conflicts and then commit the result} in the output of the \texttt{git merge} command.
In our experiments, we use open-source \github repositories described in Section~\ref{sec:evaluation_setup}.

\subsection{Feature Extraction} \label{Subsection:Feature Extraction}

To train a classifier, we need to extract potentially predictive features from merge scenarios. Our goal is to use features whose extraction is computationally inexpensive such that the prediction can be used in practice. 
We identify these features by surveying the literature on merge conflicts and related areas, such as code evolution or software maintenance. 

In Table~\ref{tab:features}, we categorize the identified features into \featureSetNum \textit{feature sets}, along with the intuition behind them, as well as any relevant related work that previously used this feature set or a variation of it. The last column in the table shows the dimension of each feature set (i.e., the number of individual values, each corresponding to a feature, used as input to the model) for the prediction task. The dimension of some of these feature sets is one, which means that they are just a scalar value. Some other feature sets have a dimension greater than one in order to represent all the needed information; such feature sets would be represented as a vector. For example, feature set \#4 is inspired from previous merging and code review studies~\cite{kononenko2018studying,fan2018early} and indicates code churn. We include this feature set since more code changes may increase the chance of conflicts. It needs 5 values to represent number of added, deleted, modified, copied, and renamed files. Feature sets \#4, \#5, \#7, and \#8 are vectors with size $5$, $2$, $12$, and $4$, respectively, and the other feature sets are scalars. In the end, each merge scenario is represented by a total of \featureNum features. We do not rely on language-specific features; all of our feature sets are language-agnostic.

\begin{table*}[!t]
\centering
\caption{Feature Sets Used For Training the Merge Conflict Predictor}
\label{tab:features}
\begin{tabularx}{\linewidth}{|l|X|l|X|l|}
    \hline
    \textbf{No.} & \textbf{Feature Set} & \textbf{References} & \textbf{Intuition for Including this Feature Set} & \textbf{Dimension} \\
    \hline
    1& No. of simultaneously changed files in two branches & \cite{sarma2012palantir,lessenich2018indicators} & The increase in simultaneously changed files increases the chance of conflicts. If the value of this feature is zero, no conflict can occur. & 1 \\ \hline
    2& No. of commits between the ancestor and the last commit in a branch & \cite{lessenich2018indicators, kononenko2018studying, fan2018early} & Having more commits means more changes in a branch, which may increase conflicts & 1 \\ \hline
    3& Commit density: No. of commits in the last week of development of a branch & \cite{lessenich2018indicators} &  Lots of recent activity may increase the chance of conflicting changes & 1 \\ \hline
    4& No. added, deleted, renamed, modified, and copied files in a branch & \cite{lessenich2018indicators, kononenko2018studying, fan2018early} & More code changes may increase the chance of conflicts & 5 \\ \hline
    5& No. added and deleted lines in a branch & \cite{lessenich2018indicators, fan2018early} &  More code changes may increase the chance of conflicts  & 2\\ \hline
    6& No. of active developers in a branch & \cite{kononenko2018studying, fan2018early, estler2014awareness} & Having more developers increases the chances of getting inconsistent changes & 1 \\ \hline
    7& The frequency of predefined keywords  in the commit messages in a branch. We use 12 key-words: fix, bug, feature, improve, document, refactor, update, add, remove, use, delete, and change. & \cite{fan2018early} &  These keywords can provide a high-level overview of the types of code changes and their purpose. Certain types of changes may be more prone to conflicts. & 12 \\ \hline
    8& The minimum, maximum, average, and median length of commit messages in the branch & \cite{fan2018early} & The length of a commit message can be an indicator of its quality & 4 \\ \hline
    9& Duration of the developement of the branch in hours & \cite{estler2014awareness} & The longer a branch exists for, the more likely it is for inconsistent changes to happen in one of the other branches & 1 \\ \hline
    \multicolumn{4}{|l|}{Total number of features:} & 28 \\ \hline
\end{tabularx}
\vspace{-0.5cm}
\end{table*}

The feature sets shown in Table~\ref{tab:features} are on different granularity levels. Feature set \#1, \textit{No. of simultaneously changed files}, is a \textit{merge-level} feature set, which means that this feature set is extracted once for a given merge scenario. All the other feature sets are \textit{branch-level}, which means that these feature sets are extracted from each branch separately. 
Each feature set should have a single value for each merge scenario. This means that we need to combine the two values of branch-level feature sets somehow. 
Since the choice of the combination operator may impact the performance of the classifier, we empirically determine the best combination operator to use, as we describe in Section~\ref{sec:evaluation_setup}.

For all the feature sets listed in the table, we use the \texttt{git log} command, with different parameters depending on the feature set, to extract their values. Our artifact page contains the exact \texttt{git log} commands we use.

\subsection{Prediction}  \label{Subsection:Prediction}

The aim of the prediction phase is to train a binary classifier that is capable of predicting whether a merge scenario is safe or conflicting after learning from the development history of a different set of merge scenarios. Merge conflict data gathered from \git history is highly imbalanced; specifically, the number of merge scenarios without conflicts is much higher than merge scenarios with conflicts. Imbalanced data prevents the standard variation of most classification methods from working well for the minor class (i.e., the class with fewer data points). There are several techniques in the field of machine learning that have been designed to overcome this problem\cite{haixiang2017learning}, including data resampling, tree-based models, an ensemble learning approach, or alternative cost functions,
Resampling methods, such as Synthetic Minority Over-sampling Technique (SMOTE)~\cite{chawla2002smote}, are computationally expensive which is why we opt for ensemble learning techniques that combine multiple simpler classification models, allowing them to handle imbalanced data.
Specifically, we use Random Forest, which is an ensemble tree-based classification method. 
We train a separate classifier for each programming language.

%% file: sections/evaluation_setup.tex
\section{Data Collection Process}
\label{sec:evaluation_setup}

%
%
%

The first step for applying our methodology from Section~\ref{sec:methodology} is to choose the target repositories to be analyzed.
Since the selected data may greatly impact our results, we dedicate this section to describe the data collection process in detail.
We then present the specific methods used to answer each RQ in Sections~\ref{sec:rq1} and~\ref{sec:rq2}.

We focus on open-source repositories in this work. We, thus, need to ensure that the selected repositories are of high quality and reflect real-world development practices. As a proxy for quality, we look for well-engineered repositories (i.e., real-world engineered software projects~\cite{munaiah2017curating}) that are also popular. Specifically, we use the following criteria:

\begin{itemize}
\item \textbf{\textit{Popularity:}} Intuitively, more active and useful repositories attract more attention, reflected in the number of stars, issues, forks, etc. Similar to previous studies\cite{accioly2018understanding,accioly2018analyzing}, we use the number of stars as a filtering criterion.
\item \textbf{\textit{Quality:}} Even though the number of stars represents some measure of quality, not all popular repositories are suitable for our study. For instance, there are a number of repositories that only consist of code examples and interview questions that are highly starred but are not suitable for studying merge conflicts since they do not represent a collaborative effort to build a software system. Hence, we apply further quality measures for our repository selection. We use reaper~\cite{munaiah2017curating} to detect well-engineered software repositories and avoid analyzing personal or toy repositories. Reaper uses various repository characteristics such as community support, continuous integration, documentation, history, issues, license, and unit testing to classify well-engineered software repositories using a random forest classier. We use reaper's released dataset~\cite{reaper} (downloaded on September 15, 2018) and select all repositories in that list that have been classified as well-engineered repositories. 
\item \textbf{\textit{Programming Language:}} We choose all \progLangNum programming languages that the reaper dataset supports: \progLangs.
\end{itemize}

Considering the three criteria mentioned above, we sort the well-engineered repositories in each programming language separately based on the number of stars. We then select the top \initreposNumperlang repositories from each language, for a total of \initreposNum repositories as the initial list. For practical limitations with respect to computational resources for replaying thousands of merge scenarios from that many repositories, we only consider repositories whose size is less than $1$ GB and focus on the latest $1,000$ merge scenarios in each repository. We focus on active repositories and therefore eliminate any moved or archived repositories from that initial list. Moreover, to avoid analyzing the same merge scenario multiple times, we only analyze the main repositories and eliminate the forked versions. After these eliminations, we are left with a total of \reposNum repositories that we use for our study. The list of the selected repositories we use in our experiments is available on our artifact page. 



\begin{figure}[t!]
\centering
  \includegraphics[width=0.45\textwidth]{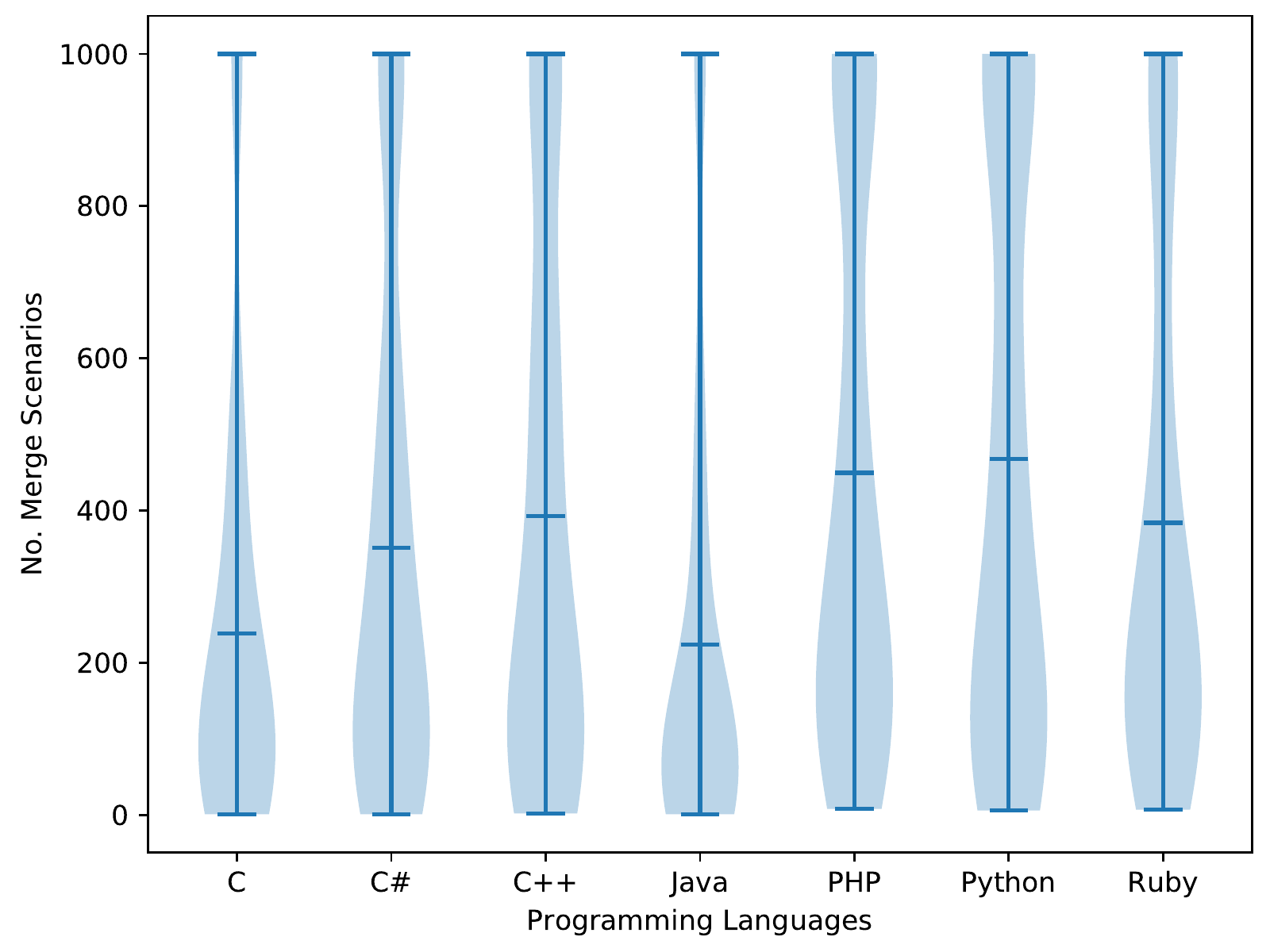}
  \caption{The Distribution of Merge Scenarios}
    \vspace{-0.5cm}
  \label{fig:merges}
\end{figure}

\begin{figure}[t!]
\centering
  \includegraphics[width=0.45\textwidth]{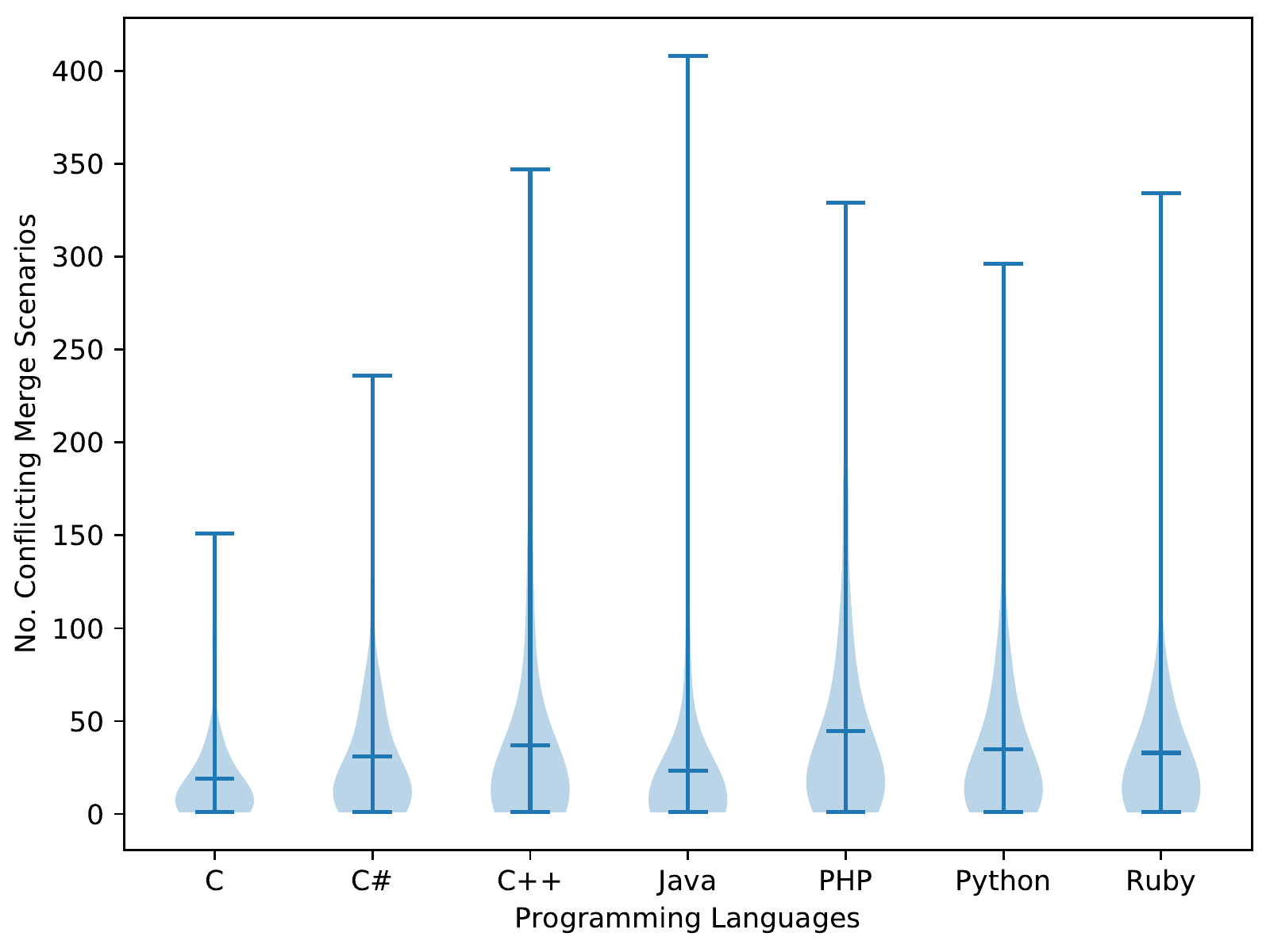}
  \caption{The Distribution of Conflicting Merge Scenarios}
  \vspace{-0.5cm}
  \label{fig:conflicts}
\end{figure}

After choosing the target repositories, we analyze their latest $1,000$ merge scenarios. We collect \mergeScenarioNum merge scenarios in total. Figure~\ref{fig:merges} illustrates the distribution of merge scenarios for repositories in different programming languages. While most of the Java repositories have less than $200$ merge scenarios, the distribution of merges in languages such as C++, PHP, and Python is close to uniform. It means that in these programming languages, we can see the repositories with a different number of merges, from zero to $1,000$ (our pre-defined threshold) with relatively the same chance. In Figure~\ref{fig:conflicts}, we show the distribution of conflicting merge scenarios in the same way. While the range of the conflicting merge scenarios is different across the languages, maximum of $150$ in C to more than $400$ in Java, the shape of their distribution is the same and the median is less than $50$. 

We show the number of repositories, merge scenarios, conflicting merge scenarios, and the conflict rate of each programming language in detail in Table~\ref{tab:mainData}. Out of \mergeScenarioNum scenarios, \mergeScenarioNumConflicting have at least one conflict in their textual files, such as code files or documentation files. In our data, the conflict rate across the different programming languages is \confRate. In such imbalanced data, we need to select and train the proper prediction models to make sure that our classifiers can perform well for correctly predicting both safe and conflicting merge scenarios.

\begin{table*}[t!]
\centering
\caption{Descriptive Statistics of Our DatasSet}
\label{tab:mainData}
\begin{tabular}{|l|l|l|l|l|}
\hline
\textbf{Programming Languages} & \textbf{\# Repositories} & \textbf{\# Merges} & \textbf{\# Conflicting Merges} & \textbf{Conflict Rate (\%)} \\ \hline
\textit{\textbf{C}}            & 80                       & 18,824             & 1,308                          & 6.95                        \\ \hline
\textit{\textbf{C++}}          & 109                      & 42,420             & 3,621                          & 8.54                        \\ \hline
\textit{\textbf{C\#}}          & 110                      & 38,945             & 3,153                          & 8.1                         \\ \hline
\textit{\textbf{Java}}         & 120                      & 36,853             & 2,190                          & 5.94                        \\ \hline
\textit{\textbf{PHP}}          & 112                      & 50,342             & 4,737                          & 9.41                        \\ \hline
\textit{\textbf{Python}}       & 106                      & 49,583             & 5,533                          & 11.16                        \\ \hline
\textit{\textbf{Ruby}}         & 106                      & 40,690             & 3,192                          & 7.84                        \\ \hline \hline
\textit{\textbf{Sum}}  & 744                      & 267,657            & 21,734                         & - \\ \hline
\textit{\textbf{Weighted Average}}  & -                      & -            & -                        & 8.12                        \\ \hline
\end{tabular}
\end{table*}

\section{RQ1: Which characteristics of merge scenarios have more impact on conflicts?} 
\label{sec:rq1}

In RQ1, we are interested in identifying which feature sets are more important for predicting conflicts.
We first describe the analysis methods we use, given the data collected in Section~\ref{sec:evaluation_setup} and then present the results.

\subsection{Method}

To answer RQ1, we analyze the \featureSetNum feature sets in Table~\ref{tab:features} to see which of them are more important. We analyze importance in two ways. (1) We calculate Spearman's rank-order correlation\cite{kendall1939distribution}, which is a non-parametric measure, between the feature sets and the existence of conflicts. This is the same correlation method used in previous work to determine the effectiveness of various features for predicting conflicts~\cite{lessenich2018indicators}. 
(2) We use decision trees to analyze the importance of each feature set, since the results of decision trees are easier to interpret than other classifiers. A decision tree aims to find a single feature set in each level based on which it can classify the data in the most optimized way. For feature sets that have more than one feature, we calculate the average of the importance of their individual features.

\subsection{Results}
\label{sec:rq1-res}

\subsubsection{Correlation-based analysis} \label{subsec:rq1}

We first analyze the Spearman's rank-order correlation between the feature sets and merge conflicts, as shown in Table~\ref{table:corr}. We calculate the correlation and the corresponding $p$-value for each feature set separately. The $p$-values of Feature Sets \#7 to \#9, for all languages, are greater than $0.05$ showing that there is no significant correlation between the conflicts and these feature sets. Following previous work~\cite{lessenich2018indicators} and statistics guidelines~\cite{anderson2012new}, we consider correlation coefficients $>= 0.6$ as strong (and highlight them in the table), $0.4 - 0.59$ as medium, and $0.2 - 0.39$ as weak. We only consider statistically significant correlations whose $p$-value $< 0.05$.

The first feature set has the highest correlation, between $0.52$ to $0.60$, for the different programming languages. However, this correlation is strong only in the case of Java and PHP. Given that the higher the number of simultaneously edited files, the higher the chances of conflicts, the high correlation matches our intuition. The second feature set, which is the number of commits that happened in each branch since they diverged, has a weak correlation but this correlation is also much higher than the remaining feature sets. This means that there are only two feature sets that have at least a weak correlation with merge conflicts. The other features are have extremely low correlation coefficients ($< 0.2$) or are insignificant ($p$-value $>= 0.05$).
 
While we do not use the same exact features from the previous work by Le{\ss}enich et al.\cite{lessenich2018indicators}, we can confirm their findings in terms of lacking correlation between \git features of merge scenarios and conflicts. This gives us confidence that the lack of correlations we find for most features is correct. However, we argue that this lack of correlation does not necessarily mean merge conflicts are not predictable, as we show later in the results of RQ2.

Although we report the correlation and importance of feature sets for different programming languages separately, it is important to note that we do not expect to see significant differences between the feature sets in different programming languages since our feature sets are language-agnostic. Our results in Table~\ref{table:corr} confirm that.

\begin{table*}[t!]
\centering
\caption{Spearman's Rank-Order correlation coefficients (CC) and the corresponding $P$-Values (p) between each feature set and the existence of conflicts, separated by language. The correlations that are equal or greater than $0.6$ with $P$-Values (p) less than $0.05$ are highlighted.}
\label{table:corr}
\begin{tabular}{|l|l|l|l|l|l|l|l|l|l|l|l|l|l|l|}
\hline
\multicolumn{1}{|c|}{\multirow{2}{*}{\textbf{\begin{tabular}[c]{@{}c@{}}Feature\\ Set\end{tabular}}}} & \multicolumn{2}{c|}{\textbf{C}} & \multicolumn{2}{c|}{\textbf{C++}} & \multicolumn{2}{c|}{\textbf{C\#}} & \multicolumn{2}{c|}{\textbf{Java}} & \multicolumn{2}{c|}{\textbf{PHP}} & \multicolumn{2}{c|}{\textbf{Python}} & \multicolumn{2}{c|}{\textbf{Ruby}} \\ \cline{2-15} 
\multicolumn{1}{|c|}{} & \multicolumn{1}{c|}{\textit{\textbf{CC}}} & \multicolumn{1}{c|}{\textit{\textbf{p}}} & \multicolumn{1}{c|}{\textit{\textbf{CC}}} & \multicolumn{1}{c|}{\textit{\textbf{p}}} & \multicolumn{1}{c|}{\textit{\textbf{CC}}} & \multicolumn{1}{c|}{\textit{\textbf{p}}} & \multicolumn{1}{c|}{\textit{\textbf{CC}}} & \multicolumn{1}{c|}{\textit{\textbf{p}}} & \multicolumn{1}{c|}{\textit{\textbf{CC}}} & \multicolumn{1}{c|}{\textit{\textbf{p}}} & \multicolumn{1}{c|}{\textit{\textbf{CC}}} & \multicolumn{1}{c|}{\textit{\textbf{p}}} & \multicolumn{1}{c|}{\textit{\textbf{CC}}} & \multicolumn{1}{c|}{\textit{\textbf{p}}} \\ \hline
\textit{\textbf{1}} & 0.53 & 0.0 & 0.58 & 0.0 & 0.52 & 0.0 & \cellcolor{gray!25}{0.60} & 0.0 & \cellcolor{gray!25}{0.60} & 0.0 & 0.53 & 0.0 & 0.58 & 0.0 \\ \hline
\textit{\textbf{2}} & 0.30 & 0.0 & 0.33 & 0.0 & 0.28 & 0.0 & 0.33 & 0.0 & 0.35 & 0.0 & 0.29 & 0.0 & 0.32 & 0.0 \\ \hline
\textit{\textbf{3}} & -0.17 & 0.0 & -0.17 & 0.0 & -0.07 & 0.0 & -0.14 & 0.0 & -0.19 & 0.0 & -0.08 & 0.0 & -0.01 & 0.0 \\ \hline
\textit{\textbf{4}} & -0.15 & 0.0 & -0.14 & 0.0 & -0.07 & 0.0 & -0.13 & 0.0 & -0.18 & 0.0 & -0.08 & 0.0 & -0.01 & 0.0 \\ \hline
\textit{\textbf{5}} & -0.14 & 0.0 & -0.13 & 0.0 & -0.07 & 0.0 & -0.14 & 0.0 & -0.19 & 0.0 & -0.08 & 0.0 & -0.10 & 0.0 \\ \hline
\textit{\textbf{6}} & -0.02 & 0.01 & 0.02 & 0.0 & 0.03 & 0.0 & 0.03 & 0.0 & 0.05 & 0.0 & 0.0 & 0.89 & 0.01 & 0.21 \\ \hline
\textit{\textbf{7}} & 0.01 & 0.13 & 0.0 & 0.95 & 0.01 & 0.25 & 0.0 & 0.57 & 0.01 & 0.22 & 0.0 & 0.58 & 0.01 & 0.22 \\ \hline
\textit{\textbf{8}} & 0.0 & 0.69 & 0.0 & 0.98 & 0.01 & 0.32 & 0.01 & 0.25 & 0.0 & 0.75 & 0.01 & 0.24 & 0.0 & 0.75 \\ \hline
\textit{\textbf{9}} & 0.0 & 0.98 & 0.0 & 0.99 & 0.0 & 0.55 & 0.0 & 0.76 & 0.0 & 0.67 & 0.01 & 0.11 & 0.0 & 0.65 \\ \hline
\end{tabular}
\end{table*}

\subsubsection{Decision-tree based analysis} 

As a different way of measuring feature importance, we use decision trees to determine the importance of our feature set for predicting conflicts in each programming language. The results are shown in Table~\ref{table:importance}, where the feature importance is a value between $0$ and $1$. Again, we find that the number of simultaneously changed files (Feature Set \#1) is the most important feature by far. The high impact of the number of simultaneously changed files can be intuitively explained since more in-parallel changes increase the likelihood of conflicts, and the chance of conflicts is zero if there are no simultaneously changed files. The number of commits (Feature Set \#2), edited files (Feature Set \#4), and active developers (Feature Set \#6) are also slightly more important than the other ones. However, apart from Feature Set \#1, all features seem to have very low importance for the classifier. 
Similar to the correlation-based analysis, we find that the importance of feature sets is relatively similar for all programming languages.

Our results suggest that commit message information (Feature Sets \#7 and \#8) is not important for detecting conflicts. Since commit messages contain information about the evolution of a repository (e.g., indications of types of code changes), we intuitively thought that they may have an impact on conflicts. However, the feature sets related to commit messages we currently extract are intentionally lightweight to keep execution time low. It is, therefore, hard to conclude if commit messages are indeed altogether useless in this context or if different types of feature sets (e.g., taking word embedding of the commit messages into account) may lead to more meaningful relationships. This finding is important since unlike the other numerical features, analyzing the commit messages is computationally expensive due to text processing.

\begin{findingenv}{RQ1}{rq1}
The No. of simultaneously changed files is the most important feature for predicting merge conflicts. The No. of commits in each branch shows a weak correlation, but a much lower importance level by the decision tree. Remaining feature sets show low correlation coefficients and importance.
\end{findingenv} 

\section{RQ2: Are merge conflicts predictable using only git features?} 
\label{sec:rq2}

In RQ2, our goal is to determine if merge conflicts can be predicted using our selected feature sets.
We first describe the classifiers we compare and then present our results.

\subsection{Method}
\label{sec:models}

We compare the performance of the decision tree and random forest classifiers we built using all \featureSetNum collected feature sets.
Additionally, we also create two baselines to compare our results against.
The following summarizes the four classifiers we compare.

\begin{itemize}
     \item \textbf{\textit{Decision Tree:}} A Decision Tree classifier is one of the simplest options for a binary classification task that is also robust to imbalanced data. We train a decision tree with all of our \featureSetNum feature sets. Note that this is the same classifier used to determine importance in RQ1.
    \item \textbf{\textit{Random Forest:}} To investigate if a more sophisticated classifier can make more use of the available features, we use Random Forest. 
    \item \textbf{\textit{Baseline \#1:}} The first baseline we compare to is a ``dummy'' classifier that randomly labels the data. If the data was balanced, the $f1{\text -}score$ of both classes would be around $0.5$, i.e. random guess. However, since the data is not balanced, the $f1{\text -}score$ of this classifier would be the same as the imbalance rate in the data for the conflicting class, i.e., \confRateWO, and would be $0.9188$ for the safe class. We expect that any other predictor should be better than this basic baseline in order to be useful in practice. 
    \item \textbf{\textit{Baseline \#2:}} Recall that in the results of RQ1 (Section~\ref{sec:rq1-res}), we found that Feature Set \#1, which is the number of simultaneously changed files in two branches, is the most important feature for the decision tree classifier. Therefore, as our second baseline, we use a decision tree classifier that uses \textit{only} Feature Set \#1 from Table~\ref{tab:features}. The goal of this baseline is to have a low-cost classifier that relies only on the most important feature. That way, we can determine if having the other features improves things, or is simply an added cost with no benefit.
   
\end{itemize}

\begin{table}[t!]
\centering
\caption{Feature Importance Based on a Decision Tree Classifier}
\label{table:importance}
\begin{tabular}{|l|l|l|l|l|l|l|l|}
\hline
\textbf{Feature Set} & \textbf{C} & \textbf{C++} & \textbf{C\#} & \textbf{Java} & \textbf{PHP} & \textbf{Python} & \textbf{Ruby} \\ \hline
\textit{\textbf{1}}  & 0.93       & 0.98         & 0.95         & 0.93          & 0.95         & 0.94            & 0.94          \\ \hline
\textit{\textbf{2}}  & 0.01       & 0.01         & 0.0          & 0.02          & 0.01         & 0.01            & 0.01          \\ \hline
\textit{\textbf{3}}  & 0.01       & 0.0          & 0.01         & 0.01          & 0.0          & 0.01            & 0.0           \\ \hline
\textit{\textbf{4}}  & 0.01       & 0.0          & 0.01         & 0.02          & 0.01         & 0.01            & 0.01          \\ \hline
\textit{\textbf{5}}  & 0.01       & 0.0          & 0.01         & 0.01          & 0.01         & 0.01            & 0.0           \\ \hline
\textit{\textbf{6}}  & 0.01       & 0.01         & 0.01         & 0.01          & 0.01         & 0.02            & 0.02          \\ \hline
\textit{\textbf{7}}  & 0.01       & 0.0          & 0.01         & 0.0           & 0.01         & 0.01            & 0.01          \\ \hline
\textit{\textbf{8}}  & 0.01       & 0.0          & 0.01         & 0.01          & 0.01         & 0.0             & 0.01          \\ \hline
\textit{\textbf{9}}  & 0.01       & 0.0          & 0.0          & 0.01          & 0.0          & 0.0             & 0.01          \\ \hline
\end{tabular}
\end{table}

\paragraph{Hyper-parameters} The main hyper-parameters of decision trees and random forests classifiers are (1) the minimum samples in leaves, (2) the minimum sample split, (3) the maximum depth, and (4) the total number of estimators (just for random forest). Determining the proper value for the number of estimators is important since a low number of weak classifiers may result in underfitting, while an unnecessarily high number may result in overfitting. The other hyper-parameters also balance the complexity of the models. Due to the importance of these hyper-parameters, we use grid-search with $10$-fold cross-validation to find the right hyper-parameters to use. The candidate values for each of these hyper-parameters are selected based on the typical values explored for this size of data. We use $\{2, 5, 10, 20, 35, 50\}$ as the choices for the minimum samples in leaves, $\{2, 3, 5, 10, 20, 35, 50, 75\}$ for minimum sample split, $\{1, 3, 5, 7, 11\}$ for the maximum depth, and $\{1, 3, 10, 50, 75, 100, 200, 300\}$ as the choices for the number of estimators. 
Our results show that the best hyper-parameter values for the minimum samples in leaves is $10$, for minimum sample split is $5$, for the maximum depth is $7$, and for the number of estimators is $75$.

\paragraph{Combination operators} Recall from Section~\ref{sec:methodology} that since some of our feature sets are extracted for \textit{each} branch, we need to use a combination operator to combine them into a single value for the whole merge scenario. To find the suitable combination operator to use, we train our predictors based on each of seven common combination operators: Minimum, Maximum, Average, Median, Norm- 1, Norm-2, and Concatenation operators. We then use grid search with $10$-fold cross-validation on all data points to determine the best combination operator. We find that Norm-$1$ is the best combination operator for all \progLangNum programming languages.

\begin{table*}[t!]
\centering
\caption{Merge Conflict Prediction Results. The highest values in each category are highlighted.}
\label{Table:results}
\begin{tabular}{|l|l|l|l|l|l|l|l|}
\hline
\multicolumn{1}{|c|}{\multirow{2}{*}{\textbf{\begin{tabular}[c]{@{}c@{}}Programming\\ Language\end{tabular}}}} & \multicolumn{1}{c|}{\multirow{2}{*}{\textbf{Classifier}}} & \multicolumn{3}{c|}{\textbf{Safe (Not Conflicting)}} & \multicolumn{3}{c|}{\textbf{Conflicting}} \\ \cline{3-8} 
\multicolumn{1}{|c|}{} & \multicolumn{1}{c|}{} & \multicolumn{1}{c|}{\textbf{$Precision_{S}$}} & \multicolumn{1}{c|}{\textbf{$Recall_{S}$}} & \multicolumn{1}{c|}{\textbf{$f1{\text -}score_{S}$}} & \multicolumn{1}{c|}{\textbf{$Precision_{C}$}} & \multicolumn{1}{c|}{\textbf{$Recall_{C}$}} & \multicolumn{1}{c|}{\textbf{$f1{\text -}score_{C}$}} \\ \hline  \hline
\multirow{3}{*}{\textbf{C}} & \textit{\textbf{Baseline \#2}} & \cellcolor{gray!25}{1.00} & 0.81 & 0.90 & 0.28 & \cellcolor{gray!25}{1.00} & 0.44 \\ \cline{2-8} 
 & \textit{\textbf{Decision Tree}} & 0.99 & 0.89 & 0.94 & 0.37 & 0.88 & 0.52 \\ \cline{2-8} 
 & \textit{\textbf{Random Forest}} & 0.98 & \cellcolor{gray!25}{0.96} & \cellcolor{gray!25}{0.97} & \cellcolor{gray!25}{0.56} & 0.72 & \cellcolor{gray!25}{0.63} \\ \hline  \hline
\multirow{3}{*}{\textbf{C++}} & \textit{\textbf{Baseline \#2}} & \cellcolor{gray!25}{1.00} & 0.82 & 0.90 & 0.34 & \cellcolor{gray!25}{0.99} & 0.51 \\ \cline{2-8} 
 & \textit{\textbf{Decision Tree}} & 0.99 & 0.88 & 0.93 & 0.41 & 0.91 & 0.57 \\ \cline{2-8} 
 & \textit{\textbf{Random Forest}} & 0.97 & \cellcolor{gray!25}{0.96} & \cellcolor{gray!25}{0.97} & \cellcolor{gray!25}{0.63} & 0.68 & \cellcolor{gray!25}{0.66} \\ \hline  \hline
\multirow{3}{*}{\textbf{C\#}} & \textit{\textbf{Baseline \#2}} & \cellcolor{gray!25}{0.99} & 0.83 & 0.90 & 0.32 & \cellcolor{gray!25}{0.92} & 0.48 \\ \cline{2-8} 
 & \textit{\textbf{Decision Tree}} & \cellcolor{gray!25}{0.99} & 0.85 & 0.92 & 0.35 & 0.90 & 0.51 \\ \cline{2-8} 
 & \textit{\textbf{Random Forest}} & 0.97 & \cellcolor{gray!25}{0.93} & \cellcolor{gray!25}{0.95} & \cellcolor{gray!25}{0.48} & 0.74 & \cellcolor{gray!25}{0.57} \\ \hline  \hline
\multirow{3}{*}{\textbf{Java}} & \textit{\textbf{Baseline \#2}} & \cellcolor{gray!25}{1.00} & 0.85 & 0.92 & 0.36 & \cellcolor{gray!25}{0.99} & 0.53 \\ \cline{2-8} 
 & \textit{\textbf{Decision Tree}} & 0.99 & 0.90 & 0.94 & 0.44 & 0.93 & 0.60 \\ \cline{2-8} 
 & \textit{\textbf{Random Forest}} & 0.98 & \cellcolor{gray!25}{0.95} & \cellcolor{gray!25}{0.97} & \cellcolor{gray!25}{0.58} & 0.83 & \cellcolor{gray!25}{0.68} \\ \hline  \hline
\multirow{3}{*}{\textbf{PHP}} & \textit{\textbf{Baseline \#2}} & \cellcolor{gray!25}{1.00} & 0.83 & 0.91 & 0.38 & \cellcolor{gray!25}{0.99} & 0.55 \\ \cline{2-8} 
 & \textit{\textbf{Decision Tree}} & 0.99 & 0.87 & 0.93 & 0.44 & 0.93 & 0.59 \\ \cline{2-8} 
 & \textit{\textbf{Random Forest}} & 0.98 & \cellcolor{gray!25}{0.93} & \cellcolor{gray!25}{0.95} & \cellcolor{gray!25}{0.54} & 0.82 & \cellcolor{gray!25}{0.65} \\ \hline  \hline
\multirow{3}{*}{\textbf{Python}} & \textit{\textbf{Baseline \#2}} & \cellcolor{gray!25}{1.00} & 0.82 & 0.90 & 0.29 & \cellcolor{gray!25}{1.00} & 0.45 \\ \cline{2-8} 
 & \textit{\textbf{Decision Tree}} & \cellcolor{gray!25}{1.00} & 0.87 & 0.93 & 0.36 & 0.95 & 0.52 \\ \cline{2-8} 
 & \textit{\textbf{Random Forest}} & 0.98 & \cellcolor{gray!25}{0.94} & \cellcolor{gray!25}{0.96} & \cellcolor{gray!25}{0.49} & 0.74 & \cellcolor{gray!25}{0.59} \\ \hline \hline
\multirow{3}{*}{\textbf{Ruby}} & \textit{\textbf{Baseline \#2}} & \cellcolor{gray!25}{1.00} & 0.84 & 0.91 & 0.33 & \cellcolor{gray!25}{1.00} & 0.50 \\ \cline{2-8} 
 & \textit{\textbf{Decision Tree}} & \cellcolor{gray!25}{1.00} & 0.89 & 0.94 & 0.41 & 0.96 & 0.57 \\ \cline{2-8} 
 & \textit{\textbf{Random Forest}} & 0.98 & \cellcolor{gray!25}{0.96} & \cellcolor{gray!25}{0.97} & \cellcolor{gray!25}{0.59} & 0.72 & \cellcolor{gray!25}{0.65} \\ \hline
\end{tabular}
\end{table*}

\paragraph{Performance measures} It is important to note that accuracy is not a good performance measure for imbalanced data since the potential influence of misclassification of conflicting merges would be much lower than safe merges.  For example, imagine that there are 100 merge scenarios with 20 of them having conflicts and 80 without conflicts. A naive classifier that simply classifies everything as not conflicting would achieve a misleading accuracy of 80\%. Hence, instead of accuracy, we use precision, recall, and $f1{\text -}score$ to evaluate the performance of the classifiers. 
 
However, reporting the recall and precision only for the conflicting class gives a partial view of how a detector would perform in practice. 
Given that safe merge scenarios occur much more often than conflicting ones, we also need to make sure that we have good recall and precision for the safe class.
Therefore, we report all our performance measures for both the conflicting (C) and safe (S) classes as follows.

For a given merge scenario, any of the binary classifiers we compare predicts either a conflict or a not conflict (i.e., a \textit{safe merge}).
After running a given classifier on all our evaluation data using $10$-fold cross-validation, we consider each class separately as the target class and calculate precision, recall, and $f1{\text -}score$ according to the following definitions:

\begin{itemize}
\item \textit{\textbf{True Positive (TP):}} The target class is labeled correctly.
\item \textit{\textbf{False Positive (FP):}} The non-target class is incorrectly labeled as the target class.
\item \textit{\textbf{True Negative (TN):}} The non-target class is labeled correctly.
\item \textit{\textbf{False Negative (FN):}} The target class is incorrectly labeled as the non-target class. 
\end{itemize}

The evaluation measures are:

\begin{equation}
precision  = \frac{TP}{TP + FP}
\end{equation}

\begin{equation}
recall  = \frac{TP}{TP + FN}
\end{equation}

\begin{equation}
f1{\text -}score  = 2  * \frac{Precision * Recall}{Precision + Recall}
\end{equation}

For example, assume the ground truth is  $\{C, S, S, C, S, S\}$, and a predictor labels the data as follows $\{S, S, C, C, S, C\}$, then $recall_{C} = 1/2 = 0.5$, $precision_{C} = 1/3 = 0.33$, $recall_{S} = 2/4 = 0.5$, $precision_{S} = 2/3 = 0.67$.

%% file: sections/results.tex
\subsection{Results}

%
%
%
%

Table \ref{Table:results} shows our results for RQ2. Note that we do not show the results of Baseline \#1 since it can be calculated based on the bias in the data and serves as a minimum threshold that any useful predictor needs to achieve. 

\subsubsection{Decision Trees vs. Baseline \#2} We first compare Baseline \#2, which is a simple decision tree that uses the most important feature determined in Section~\ref{sec:rq1-res}, to the Decision Tree classifier that uses all features. Table~\ref{Table:results} shows that the Decision Tree classifier that uses all features achieves a higher $f1{\text -}score$ for both classes when compared to Baseline \#2. This suggests that despite Feature set \#1 being the most important feature, adding the other features to the classifier does improve the results. 

Additionally, both the Decision Tree classifier and Baseline \#2 exceed the performance of Baseline \#1, which is a ``dummy'' classifier that randomly labels the data by considering the imbalance rate. This shows that there is gained value in designing a ``real'' classifier.

\subsubsection{Decision Trees vs. Random Forest} Given that the Decision Tree classifier with all features outperforms Baseline \#2, we now compare the Decision Tree classifier to Random Forest to determine if a more sophisticated classifier can achieve better results. The results in Table~\ref{Table:results} show that the Random Forest classifier achieves the highest $f1{\text -}score$ for both safe and conflicting merges. This shows that using all features along with a more advanced ensemble machine learning classifier does indeed achieve better results. Another observation is that all the classifiers seem to perform consistently across the different programming languages.

\subsubsection{Conflicting class} We now focus on the Random Forest classifier and discuss the results for the conflicting class in more detail. The table shows that $recall_{C}$ ranges from \rCMin to \rCMax for the different programming languages. This means that the predictor can correctly identify most of the conflicting merge scenarios. The table shows that $precision_{C}$ is in a lower range, varying from \pCMin to \pCMax. Overall, the $f1{\text -}score_{C}$ ranges from \fCMin to \fCMax across the \progLangNum languages. 

\subsubsection{Safe class} In terms of not conflicting, or safe, merge scenarios, Table~\ref{Table:results} shows that Random Forest's $recall_{S}$ is between \rNCMin to \rNCMax. This is a high recall rate and means that the predictor is able to correctly identify most of the automatically mergeable merge scenarios (i.e., those that will not result in conflicts). The precision of this class is between \pNCMin to \pNCMax for different programming languages, meaning that almost all of the merge scenarios that are predicted as safe are actually safe. Overall, the $f1{\text -}score_{NC}$ ranges between \fNCMin to \fNCMax for the different programming languages.

\begin{findingenv}{RQ2}{rq2}
We find that a Random Forest classifier based on light-weight \git features can successfully predict conflicts for different programming languages. However, the $f1{\text -}score$ of the safe class is much higher than the conflicting class.
\end{findingenv}

We, finally, note that the average time for predicting the status of a given merge scenario, including the feature extraction process, is $\predictionTime\pm\predictionTimeSTD$.
This makes our predictor fast enough to use in practice.

\section{Implications and Discussion}
\label{sec:implications}

We now discuss what our prediction results may mean for avoiding complex merge conflicts in practice.

The recall of merge conflicts is relatively high (\rCMin to \rCMax), which means that the classifier can identify an acceptable portion of conflicts, if it is used as a replacement of speculative merging altogether. 
Notifying developers of these potential conflicts would allow them to merge early and avoid the conflict becoming more complex.
The downside is that the precision of predicting conflicts is lower (\pCMin to \pCMax), which means that developers may perform a merge earlier than needed (i.e., perform a merge when there is no conflict to resolve).
In practice, this may not be a big problem since frequent merges are encouraged to avoid conflicts in the long term.
 
However, instead of completely replacing speculative merging and running the risk of false positive notifications to developers, we advocate for using a merge-conflict predictor as a pre-filtering step for speculative merging~\cite{brun2013early,brun2011proactive} or continuous merging~\cite{guimaraes2012improving} in developers' work environments (e.g., their IDE).
Both recall and precision of our classifier for safe merges are considerably high (recall between \rNCMin to \rNCMax and precision in the range of \pNCMin to \pNCMax). The precision of safe merge scenarios in the context of pre-filtering them out from speculative merging is important, since we want to make sure that eliminated merge scenarios are actually safe. Given the high precision and the fact that conflict rates are typically low (\confRate), this means that a subsequent proactive conflict detection tool will accurately eliminate a large number of safe merge scenarios from its analysis, thus potentially saving costs.

%% file: sections/threats.tex
\section{Threats to Validity}
In this section, we discuss some of the potential threats to the validity of our study.

\subsection{Internal Validity}
\texttt{git merge} can use several merging algorithms, and the choice of algorithm used may impact the results. We employ the default one (recursive merging strategy) since developers typically do not change the default configuration of Git merge.
 
Rebasing is another strategy for integrating changes from different branches.
When \texttt{git rebase} is used instead of \texttt{git merge} or when the \texttt{--rebase} option is used while pulling, a linear history is created and no explicit merge commits will exist.
Therefore, there is a chance that we miss some merge scenarios since we detect merge scenarios based on the number of parents of a commit. Unfortunately, there is no precise methodology to extract rebased merge scenarios since there is no information in \git about them.

We eliminate $n$-way (octopus) merging and only focus on $3$-way merging where each merge commit has exactly two parents. This may eliminate some merge scenarios. However, $3$-way merging happens more often in practice.

We use a set of candidate values for the hyper-parameters we use for our classifiers and find the best option by using a grid search. We created these candidate values based on our intuition and the heuristics in the literature about the hyper-parameters for machine learning techniques. However, we cannot guarantee that we found the globally optimal values for our hyper-parameters.

We do not consider the chronological order or timeline of commits in any way. In other words, we do not train our model with a subset of merge scenarios and test them only with the subsequent ones. While such \textit{time travel} is often a threat in prediction studies, we believe that the impact is low in our context since most of the features we use in our prediction model are not time sensitive. For example, the number of co-modified files or the number of changed lines is not dependent on the time in a project. In contrast, features such as the name of the file or code component being modified (which we do not use in our work) are time sensitive since they may change significantly over time in a project.

\subsection{External Validity}
While we have a large-scale empirical study, our evaluation is still limited to \reposNum open-source repositories in \github in \progLangNum popular programming languages. Our results may not address merge conflict prediction in other programming languages. However, our work is, to the best of our knowledge, the largest study for merge conflict prediction, to date, that also studies multiple programming languages.
While we need to train a separate predictor for repositories in each programming language, this does not have a negative impact on proactive conflict detection in practice since the language of each repository is known beforehand and the appropriate classifier can be used.

%% file: sections/conclusion.tex
\section{Conclusion}

In this paper, we investigated whether predicting merge conflicts is feasible, with the long-term motivation of using it in the context of proactive conflict detection.
We extracted \mergeScenarioNum merge scenarios from \reposNum repositories, written in \progLangNum programming languages, and used \featureNum light-weight features from \git to design a classifier for merge conflicts. 
We compare a Random Forest classifier to two variations of a Decision Tree classifier.
While similar to previous work, we could not find a correlation between our feature sets and conflicts, we were able to successfully design a classifier for merge conflicts.
This shows that lack of correlation does not necessarily mean that prediction is not possible.
Our results show that a Random Forest classifier with all our selected features outperformed the other classifiers we compared to.
Our high precision (\pNCMin to \pNCMax) for detecting safe merge scenarios ensures that we can eliminate merge scenarios that are labeled as safe from the speculative merging process.

As future work, we plan to investigate the characteristics of conflicts in different domains to determine if the application context can have any impact on merge conflicts. Moreover, we want to integrate our conflict predictor with speculative merging in developer's IDEs.

%% file: sections/acknowledgement.tex
\section{Acknowledgement}
\label{sec:acknowledgement}

This project has been partially funded through a 2017 Samsung  Global  Research Outreach (GRO) program.